\documentclass[iop,apjl]{emulateapj}
\usepackage{graphicx}
\usepackage{color}

\newcommand{\V}[1]{\mathbf{#1}} 
\newcommand{\T}[1]{\texttt{#1}} 

\newcommand\Alfvenic{Alfv\'enic }

\newcommand{\eqref}[1]{Eq.~(\ref{#1})}

\begin{document}

\title{Measuring Collisionless Damping in Heliospheric Plasmas using
  Field-Particle Correlations}

\author{K.~G. Klein$^{1,2}$,
and G.~G. Howes$^3$}

\affiliation{$^1$Department of Climate and Space Sciences and
  Engineering, University of Michigan, Ann Arbor, MI 48109,
  USA\\ $^2$Space Science Center, University of New Hampshire, Durham,
  New Hampshire 03824, USA\\ $^2$Department of Physics and Astronomy,
  University of Iowa, Iowa City, IA 52242, USA}

\begin{abstract}
An innovative field-particle correlation technique is proposed that
uses single-point measurements of the electromagnetic fields and
particle velocity distribution functions to investigate the net
transfer of energy from fields to particles associated with the
collisionless damping of turbulent fluctuations in weakly collisional
plasmas, such as the solar wind.  In addition to providing a direct
estimate of the local rate of energy transfer between fields and
particles, it provides vital new information about the distribution of
that energy transfer in velocity space. This
  velocity-space signature can potentially be used to identify the
  dominant collisionless mechanism responsible for the damping of
  turbulent fluctuations in the solar wind.  The application of this
novel field-particle correlation technique is illustrated using the
simplified case of the Landau damping of Langmuir waves in an
electrostatic 1D-1V Vlasov-Poisson plasma, showing that the procedure
both estimates the local rate of energy transfer from the
electrostatic field to the electrons and indicates the resonant nature
of this interaction. Modifications of the technique to enable
single-point spacecraft measurements of fields and particles to
diagnose the collisionless damping of turbulent fluctuations in the
solar wind are discussed, yielding a method with the potential to
transform our ability to maximize the scientific return from current
and upcoming spacecraft missions, such as the \emph{Magnetospheric
  Multiscale} (\emph{MMS}) and \emph{Solar Probe Plus} missions.
\end{abstract}



\maketitle 


\section{Introduction}

A grand challenge problem at the forefront of
space physics and astrophysics is to understand how the energy of
turbulent plasma flows and electromagnetic fields is converted into
energy of the plasma particles, either as heat or some other form of
particle energization. Under the typically low-density and
high-temperature conditions of turbulent plasmas in the heliosphere,
such as the solar wind, the turbulent dynamics is weakly collisional,
requiring the application of six-dimensional (3D-3V) kinetic plasma
theory to follow the evolution of the turbulence, where the damping of
the turbulent fluctuations occurs due to collisionless interactions
between the electromagnetic fields and the individual plasma
particles. Although \emph{in situ} spacecraft measurements in the
solar wind provide detailed information about the electromagnetic and
plasma fluctuations, these measurements are typically limited to one point
(or, at most, a few points) in space. Of great benefit to plasma
turbulence research would be a scheme to use single-point measurements
of the electromagnetic fields and particle velocity distribution
functions (VDFs) to diagnose the collisionless damping of the
turbulent fluctuations and to characterize how the damped turbulent
energy is distributed to particles with different velocities.

Here we present an innovative technique to identify and characterize
the collisionless mechanisms that govern the net transfer of energy
from the electromagnetic fields to the plasma particles by correlating
measurements of the electromagnetic fields and particle VDFs at a
single point in space. These \emph{field-particle correlations} yield
a local estimate of the rate of particle heating, and further provide
a characteristic \emph{velocity-space signature} of the collisionless
damping mechanism that leads to the energization of the plasma
particles.

Early attempts to explore wave-particle interactions using spacecraft
measurements sought the spatial or temporal coincidence of wave fields
with enhanced particle fluxes
\citep{Gough:1981,Park:1981,Kimura:1983}.  Later, wave-particle
correlators were flown on rockets and spacecraft to identify the
phase-bunching of electrons by correlating the counts of electrons in
a single energy and angle bin with the phase of the dominant wave
\citep{Ergun:1991a,Ergun:1991b,Muschietti:1994,Watkins:1996,Ergun:1998,Ergun:2001,Kletzing:2005,Kletzing:2006}. Motivated
by modern particle instrumentation with improved temporal and
phase-space resolution, the field-particle correlation technique
described here takes a significant leap forward by recovering the
correlation as a function of particle velocity, generating a much more
detailed velocity-space signature of the collisionless interactions.

Although the novel field-particle correlation technique devised here
is intended for use in diagnosing the damping of turbulent
fluctuations in the weakly collisional solar wind, to illustrate the
concept in a simplified framework, we present here its application to
the 1D-1V Vlasov-Poisson system to explore the collisionless damping
of electrostatic fluctuations in an unmagnetized plasma. After this
demonstration of the fundamental concept of using field-particle
correlations to investigate collisionless damping of fluctuations, we
discuss the application of this technique to spacecraft observations
of solar wind turbulence.

\section{Particle Energization in a Vlasov-Poisson Plasma}

The dynamics of electrostatic fluctuations in a collisionless plasma is
governed by the Vlasov-Poisson equations, where the Vlasov equation
determines the collisionless evolution of the distribution function
for each species $s$, $f_s(x,v,t)$, and the Poisson equation
determines the self-consistent evolution of the electric field, $E(x,t)
= -\partial \phi(x,t)/\partial x$, dictated by the fluctuating charge
density in the plasma.

To diagnose the collisionless transfer of energy between fields and
particles, we define the \emph{phase-space energy density} for a
particle species $s$ as $w_s(x,v,t) = m_s v^2 f_s(x,v,t)/2$, the
energy density per unit length per unit velocity. Integrating $w_s$
over velocity yields the standard \emph{spatial energy density}, and
integrating over volume produces the total microscopic kinetic energy
of the species, $W_s$.  Splitting $f_s$ into equilibrium and perturbed
components, $f_s(x,v,t)= f_{s0}(v) + \delta f_s(x,v,t)$---where the
magnitude of $\delta f_s$ is limited only by the physical constraint
$f_s\ge 0$---we can use the Vlasov equation to obtain an equation for
the rate of change of $w_s$,
\begin{eqnarray}
\nonumber
  \frac{\partial w_s(x,v,t)}{\partial t} & = - \frac{m_sv^3}{2}\frac{\partial \delta f_s(x,v,t)}{\partial x} - \frac{q_sv^2}{2}
  \frac{\partial f_{s0}(v)}{\partial v} E(x,t) \\  & - \frac{q_sv^2}{2}
  \frac{\partial \delta f_{s}(x,v,t)}{\partial v} E(x,t).
\label{eq:dwxvdt}
\end{eqnarray}
The rate of change of $w_s$ is governed by three terms: from left to
right, the (linear) ballistic term, the linear wave-particle
interaction term, and the nonlinear wave-particle interaction term.
When integrated over space using either periodic or infinitely distant
boundary conditions, the ballistic and linear wave-particle
interaction terms yield zero net energy transfer. Only the nonlinear
wave-particle interaction term produces a net change in particle
energy.  Therefore, the term $-q_sv^2 (\partial \delta f_{s}/\partial
v) E/2$ governs the net rate of energy transfer between the
electromagnetic fields and plasma particles that is associated with
collisionless damping \citep{Howes:2016prep}.

Taking the average of \eqref{eq:dwxvdt} over the entire spatial domain---the approach taken
in quasilinear theory---provides a rigorous approach to determine the
net transfer of energy between the fields and particles, but the
spatial information necessary to perform this average is not
observationally accessible using single-point measurements.  At a
single point $x_0$, all three terms of \eqref{eq:dwxvdt} are nonzero.
These terms describe both the \emph{oscillatory energy transfer}
associated with wave motion and the \emph{secular energy transfer}
associated with a net transfer of energy between fields and particles.
Unless the collisionless damping rate is particularly strong, the
magnitude of the oscillatory energy transfer described by these terms
is generally much larger than that of the secular energy transfer, so
the key challenge is to devise a procedure to isolate the
small-amplitude rate of secular energy transfer governed by the
nonlinear wave-particle interaction term.

Note that this local approach is valuable even in numerical
simulations where full spatial information is accessible, because
there is significant evidence that energy dissipation is often
highly localized in space
\citep{Wan:2012,Karimabadi:2013,TenBarge:2013a,Wu:2013a,Zhdankin:2013,Zhdankin:2015a},
so spatial averaging may obscure the details of the local dissipation
mechanism, making it more difficult to identify the physical mechanism
responsible.

\section{Field-Particle Correlation}

The form of the nonlinear
wave-particle interaction term in \eqref{eq:dwxvdt} suggests that the
rate of change of phase-space energy density can be estimated by
correlating single-point measurements of the electric field and
particle VDFs. Below we specify the procedure to isolate the local
secular transfer of energy associated with the collisionless damping of
electrostatic fluctuations in a 1D-1V Vlasov-Poisson plasma.

Labeling discrete measurement times as $t_j \equiv j\Delta t$ for
$j=0,1,2,\ldots$, we define the single-point measurements at position
$x_0$ and time $t_j$ of the field as $E_j \equiv E(x_0,t_j)$ and the
perturbed distribution function as $\delta f_{sj}(v) \equiv \delta
f_{s} (x_0,v,t_j)$. For a correlation interval of $\tau=N\Delta t$, we
define the field-particle correlation at time $t_i$ at position $x_0$ by 
\begin{equation}
  C_1(x_0,v,t_i, \tau)\equiv \frac{1}{N}\sum_{j=i}^{i+N}-q_s\frac{v^2}{2}
  \frac{\partial\delta f_{sj}(v)}{\partial v}E_j.
  \label{eq:cfp_sum}
\end{equation}
Note that this correlation is not normalized since the product
directly corresponds to the rate of energy transfer, so the amplitude
variation of this product as a function of velocity yields valuable
information about the nature of the collisionless field-particle
interaction.

For single-point measurements, the general idea of diagnosing the
energy transfer at each point in phase space reduces to determining
the distribution of the energy transfer rate in velocity space,
producing a velocity-space signature characteristic of the physical
mechanism.  Different collisionless mechanisms are likely to have
distinct velocity-space signatures of the energy transfer.  We
illustrate this field-particle correlation analysis method for the
case of the Landau damping of Langmuir waves in a 1D-1V Vlasov-Poisson
plasma, but \emph{the concept of using field-particle correlations to
  diagnose collisionless energy transfer is extremely general}.  In
principle, this method can use single-point spacecraft measurements to
examine the energization of particles in any weakly collisional
heliospheric plasma.

\section{Numerical Results}

Using the Nonlinear Vlasov-Poisson Simulation Code \T{VP}
\citep{Howes:2016prep}, we apply field-particle correlations to
examine collisionless damping in three cases: (I) a moderately damped
standing Langmuir wave pattern with $k \lambda_{de}=0.5$; (II) a
weakly damped standing Langmuir wave pattern with $k
\lambda_{de}=0.25$; and (III) a moderately damped single propagating
Langmuir wave mode with $k \lambda_{de}=0.5$, where
  $\lambda_{de}=\sqrt{k_BT_e/4 \pi n_e q^2}$ is the electron Debye
  length.  For cases I \& II, $\delta f_e(t=0)$ is a sine wave with
wavelength $k \lambda_{de}$; for case III, $\delta f_e(t=0)$ satisfies
the Langmuir wave linear dispersion relation.

The \T{VP} code evolves the nonlinear Vlasov-Poisson
  system of equations for ion and electron species using second-order
centered finite differencing for spatial and velocity derivatives and
a third-order Adams-Bashforth scheme in time. Spatial boundary
conditions are periodic and a Green's function solution is used to
determine $\phi$. All cases have plasma parameters $T_i/T_e=1$ and
$m_i/m_e=100$ and numerical resolution $n_x=128$ and $n_v=256$ with a
simulation domain of length $L=2 \pi/k$. The cases with $k
\lambda_{de} = 0.5(0.25)$ have a resonant velocity of $\omega/k = 2.86
(4.4)v_{te}$ and a linear damping rate of $1.59\times 10^{-1}
(2.05\times 10^{-3})\omega_{pe}$.


\begin{figure}
\resizebox{3.4in}{!}{\includegraphics*[0.85in,0.75in][5.25in,6.05in]
{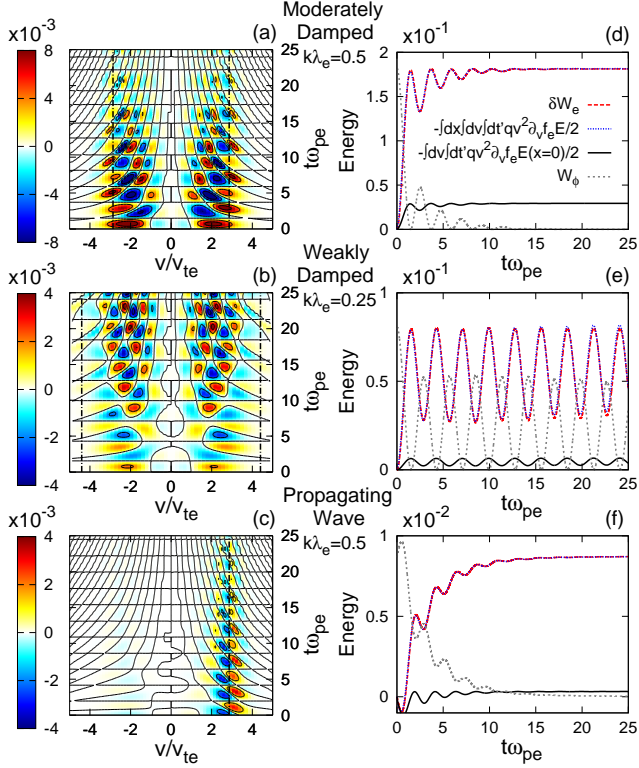}
}
\caption{ \label{fig:Energy} Rate of energy transfer in velocity space
  at $x=0$ between the fields and electrons for the cases I (a), II
  (b), and III (c) as a function of velocity $v/v_{te}$ and time $t
  \omega_{pe}$. Positive (negative) rates signify transfer
    to (from) the particle distribution. Evolution of field energy $W_\phi$
  (long-dashed gray), perturbed electron energy $\delta W_e$ (dashed
  red), phase-space and time integrated energy transfer rate (dotted
  blue), and the velocity and time integrated single-point energy
  transfer rate (solid black) corresponding to each case (d)--(f).  }
\end{figure}

In Fig.~\ref{fig:Energy}, we plot the instantaneous rate of change of
$w_e$ due to the nonlinear wave-particle interaction term, $-q_e v^2
\partial_v \delta f_e E/2$, at $x=0$ for the three cases
(a)--(c). Without calculating the correlation $C_1$ over an
appropriate time interval $\tau$, the largest rates of energy transfer
do not necessarily correspond to the resonant velocities, $v=\omega/k$
(dot-dashed black lines).  The reason is that the larger amplitude
oscillating energy transfer of the Langmuir waves masks the smaller
amplitude secular energy transfer of the collisionless damping.

Also plotted in Fig.~\ref{fig:Energy} is the time evolution of the
electrostatic field energy $W_\phi=\int dx \ E^2/8\pi$ (long-dashed
gray) and the perturbed electron energy $\delta W_e=\int dx \int dv
\ m_e v^2 \delta f_e/2$ (dashed red), showing that $\partial \delta
W_e/\partial t \simeq -\partial W_\phi/\partial t$ because the Landau
damping of Langmuir waves transfers little of the electrostatic field
energy to the ions for $m_i/m_e=100$. Thus, we focus here strictly on
energy transferred to electrons. We also plot the nonlinear
wave-particle interaction term integrated over all phase-space and
time, $-\int_0^t dt'\int dx \int dv \ q_e v^2 (\partial \delta
f_e(x,v,t')/\partial v)E(x,t')/2 $ (dotted blue), demonstrating that
this term alone contains all of the net energy transfer to the
electrons. Finally, at the single-point $x=0$, we plot the
time-integrated transfer rate, $-\int_0^t dt'\int dv q_e v^2
\partial_v \delta f_e(0,v,t') E(0,t')/2$ (solid black), showing that
we obtain a significant net transfer of energy from the field to the
electrons for both moderately damped cases with $k \lambda_{de}=0.5$.

\begin{figure}
\resizebox{3.4in}{!}{\includegraphics*[0.85in,.7in][5.65in,3.5in]
{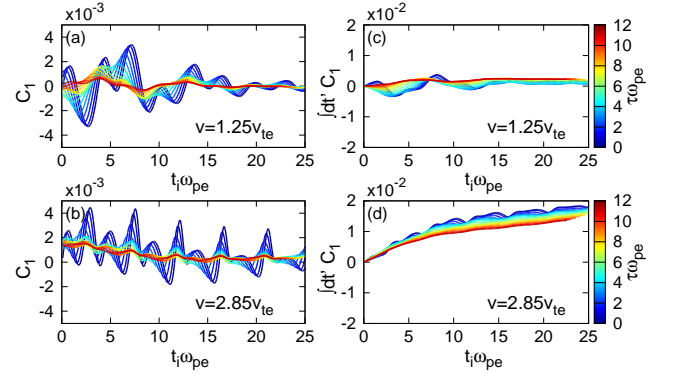}
}
\caption{ \label{fig:corr} Field-particle
  correlation~\eqref{eq:cfp_sum} at $x=0$ for case I with varying
  correlation interval $\tau$ (colorbar) for (a) off-resonant and (b)
  on-resonant velocities, along with the corresponding (c)
  off-resonant and (d) on-resonant time-integrated energy transfer
  rates $\int_0^t dt'C_1$.}
\end{figure}

To isolate the small-amplitude secular energy transfer in the presence
of a much larger amplitude oscillating energy transfer, we must select
an appropriate correlation interval $\tau$.  In
Fig.~\ref{fig:corr}, we plot the correlation $C_1(v_0,t,\tau)$ from
Eq.~\ref{eq:cfp_sum} for a range of correlation intervals
$0 \le \omega_{pe} \tau \le 12$ (colorbar) for case I both for (a) an
off-resonance velocity $v_0=1.25 v_{te}$ and (b) an on-resonance
velocity $v_0=2.85 v_{te}$. The $\tau=0$ curve (dark blue) corresponds
to a vertical slice along Fig.~\ref{fig:Energy}(a) at the selected
velocity $v_0$.  As the correlation interval $\tau$
increases, the large amplitude signal of the oscillating energy
transfer is increasingly averaged out.  For this case, the normalized
wave period is $T \omega_{pe}=4.39$, and we find that for
 correlation intervals $\tau>T$, the large-amplitude oscillating
energy transfer rate is significantly reduced, revealing the smaller
amplitude secular energy transfer rate beneath. Integrating the
correlation in time, $\int_0^tdt'\ C_1(v_0,t',\tau)$, we find (c)
little net energy at the non-resonant velocity, and (d) significant
particle energization at the resonant velocity $v_0=2.85 v_{te}$.

\begin{figure*}[ht]
\resizebox{7.1in}{!}{\includegraphics*[0.85in,0.7in][8.55in,4.6in]
{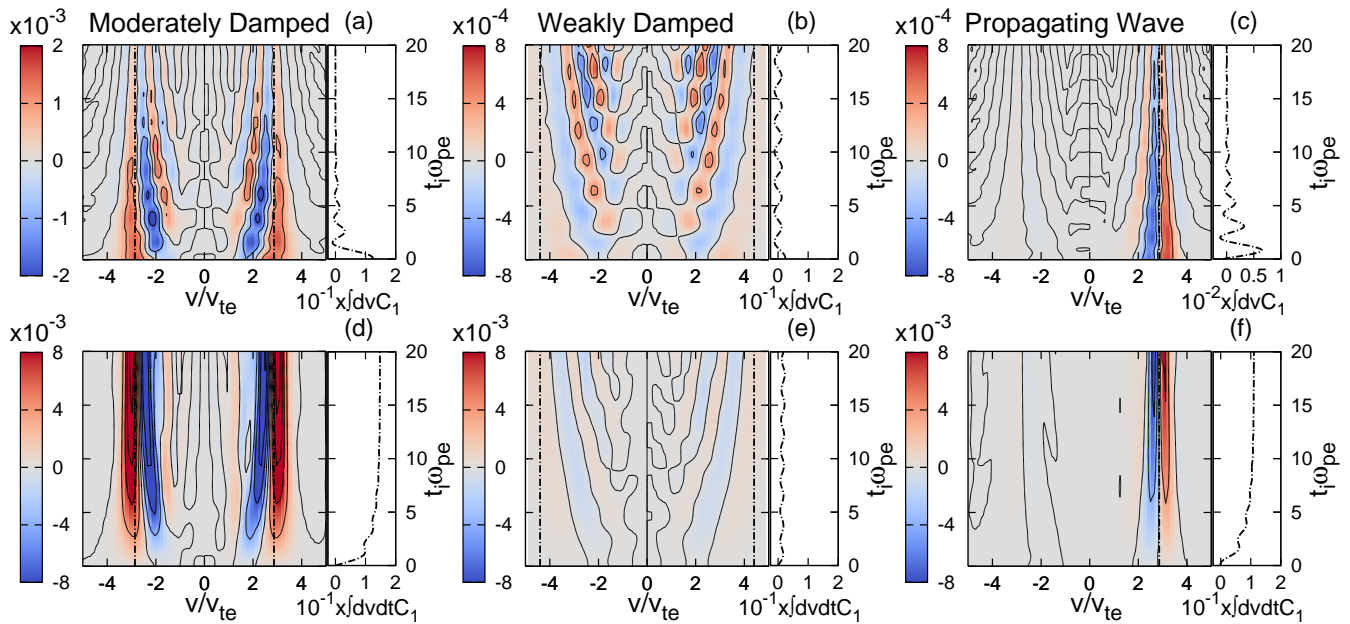}
}
\caption{ \label{fig:fullCorr} Velocity space structure of the
  field-particle correlation $C_1$ (top row) and $\int_0^t dt'C_1$
  (bottom) as well as the velocity-integration of these quantities
  (offsets) for case I, panels (a) and (d), II, (b) and (e), and III,
  (c) and (f). The correlation interval $\tau \omega_{pe}$ is set to
  $6.28$.}
\end{figure*}

In Fig.~\ref{fig:fullCorr}, we plot the key results of this Letter,
the field-particle correlations $C_1$ for $\tau \omega_{pe}=6.28$ as a
function of velocity and time for cases I--III, (a)--(c). With a
suitably long correlation interval $\tau>T$, the large
amplitude signal of the oscillating energy transfer, dominating
Fig.~\ref{fig:Energy}, is diminished, revealing the secular transfer
of energy.  This velocity-space signature of the secular energy
transfer rate is concentrated around the resonant velocity for the
moderately damped cases, indicating a resonant process.  Integrating
$C_1$ over velocity yields the net energy transfer rate at that point
in space (offset panels), equal to $jE$. This velocity integration
demonstrates a net transfer of energy to the particles, but loses all
velocity-space information that can be used to identify the nature of
the collisionless energy transfer mechanism.  The weakly damped case
has a relatively insignificant energy transfer rate.  In panels
(d)--(f), we plot the accumulated change in the electron phase-space
energy density, $\Delta w_e(x_0,v,t)=\int_0^tdt'\ C_1(v,t',\tau)$,
showing a loss of energy at $v < \omega /k$ and gain of energy at $v >
\omega /k$ for the moderately damped cases. This velocity-space
signature corresponds physically to a flattening of the distribution
function at the resonant velocity, consistent with the evolution of
the spatially averaged electron VDF predicted by quasilinear theory
\citep{Howes:2016prep}.  The nearly-monotonic increase of $\int dv
dt' C_1$ for the moderately damped cases shows that $C_1$ serves as a
measure of collisionless damping rate and not merely the presence of
monochromatic waves.

\section{Application to Solar Wind Turbulence}

Proposed collisionless
damping mechanisms in the solar wind fall into three classes: (i)
coherent collisionless wave-particle interactions, such as Landau
damping, transit-time damping, or cyclotron damping
\citep{Landau:1946,Barnes:1966,Leamon:1998b,Quataert:1998,Leamon:1999,
  Quataert:1999,Leamon:2000,Howes:2008b,Schekochihin:2009,TenBarge:2013a}
(ii) incoherent collisionless wave-particle interactions, primarily
leading to stochastic ion heating
\citep{Chen:2001,White:2002,Voitenko:2004,Bourouaine:2008,Chandran:2010a,Chandran:2010b,Chandran:2011,Bourouaine:2013};
and (iii) dissipation in coherent structures, specifically current
sheets, generally involving collisionless magnetic reconnection
\citep{Dmitruk:2004,Markovskii:2011,Matthaeus:2011,Osman:2011,Servidio:2011a,Osman:2012a,Osman:2012b,Wan:2012,Zhdankin:2013,Karimabadi:2013,Osman:2014a,Osman:2014b,Zhdankin:2015a}.
Under weakly collisional conditions, all of these mechanisms are
mediated by interactions between the electromagnetic fields and the
individual plasma particles, and therefore all  will lead to a correlation
between the fields and particle VDFs. Each mechanism is likely to
generate a distinct velocity-space signature that can be diagnosed
using the general approach of field-particle correlations.

For the case of the damping of solar wind turbulence, the appropriate
form of the correlation will depend on the specific mechanism. For
example, ion transit-time damping
\citep{Barnes:1966,Quataert:1999}---the magnetic analogue of Landau
damping---will involve a correlation of the parallel perturbed
magnetic field $\delta B_\parallel$ and the ion parallel VDF $\delta
f_i(v_\parallel)$. In addition, the appropriate component of the field
may be difficult to measure in space, such as the parallel component
of the electric field, $E_\parallel$, that leads to Landau damping.
In this case, since the electromagnetic components are related by
Maxwell's equations, another field component may be used as a proxy
(since, at least in some instances, the fields have been shown to
satisfy linear eigenfunction relationships
\citep{Salem:2012,Howes:2012a,Klein:2012,Chen:2013a}). Although the
proxy correlation no longer corresponds directly to the transfer rate
of phase-space energy density, it may nonetheless indicate the order
of magnitude of the net energy transfer and its velocity-space
signature may reveal the resonant nature of the interaction.  

The super-\Alfvenic flow of the solar wind is often
  exploited to interpret the temporal fluctuations measured by the
  spacecraft as the result of spatial fluctuations being swept past
  the spacecraft by the solar wind flow, an approximation known as the
  Taylor hypothesis \citep{Taylor:1938}. How does this solar wind flow
  impact the field-particle correlation technique? The key step is to
  perform the correlation over a suitably long correlation interval
  $\tau$ in order to average out the generally larger-amplitude
  oscillating energy transfer. Fundamentally, to average out the
  oscillatory component, all that is necessary is that the
  measurements span more than $2 \pi$ of the wave phase $\alpha$, a
  function of time and position, $\alpha(x,t)=kx - \omega t$. If the
  point of measurement is moving in space, $x_0(t)$, then the method
  simply requires that the phase $\alpha(t)=k x_0(t)- \omega t$ span
  more than $2 \pi$ over the correlation interval $\tau$, so the
  technique is essentially insensitive to the solar wind flow.
The confirmation of this
  assertion in a fully turbulent system is the focus of ongoing work.

The broadband nature of turbulent fluctuations could potentially smear
out the velocity-space signature associated with damping at a
particular wavelength. Preliminary studies
\citep{Howes:2016prep} indicate that the narrow
range of length scales over which certain damping mechanisms operate may
alleviate this potential problem. 

\begin{figure}
\begin{center}
\resizebox{2.9in}{!}{\includegraphics*[0.85in,0.7in][3.5in,4.15in]
 {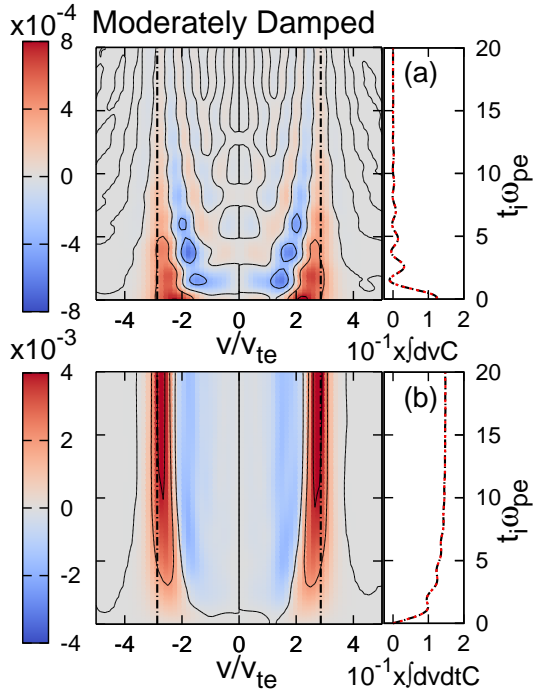}
}
\caption{ \label{fig:C2} Velocity space structure of $C_2$, top
  panel, and $\int_0^t dt' C_2$, bottom, for case I, which may serve as
  an alternative observable to $C_1$. Integration over velocity of
  $C_1$, black line in offset, and $C_2$, red line, are in agreement.}
\end{center}
\end{figure}

Finally, it may be impractical to
compute the velocity derivative of the perturbed distribution function
due to the noise and limited resolution of spacecraft measurements, so
we may use an alternative correlation
\begin{equation}
  C_2(x_0,v,t_i, \tau)= \frac{1}{N}\sum_{j=i}^{i+N} q_sv \delta f_{sj}(v)
  E_j.
  \label{eq:cfp2_sum}
\end{equation}
Note that this correlation is related to $C_1$ by an integration by
parts in velocity, so the velocity-integrated energy transfer rate is
identical to that of $C_1$ (see offset panels in Fig.~\ref{fig:C2}.
Both $C_2(v,t,\tau)$ and time-integrated correlation $\int_0^t
C_2(v,t', \tau) dt'$ with $\tau \omega_{pe}=6.28$ for case I are
plotted in Fig.~\ref{fig:C2}, yielding a velocity-space signature that
indeed indicates a resonant process.

\section{Conclusion} 

Here we present a novel field-particle correlation technique that
requires only single-point measurements of the electromagnetic fields
and particle VDFs to return an estimate of the net rate of energy
transfer between fields and particles.  Furthermore, this innovative
method yields valuable information about the distribution of this
energy transfer in velocity space, providing a vital new means to
identify the dominant collisionless mechanisms governing the damping
of the turbulent fluctuations beyond that provided by measurements of
velocity-integrated quantities such as $\V{j} \cdot \V{E}$.

This field-particle correlation technique fully exploits the vast
treasure of information contained in the \emph{fluctuations} of the
particle VDFs, potentially enabling new discoveries using single-point
spacecraft measurements.  We believe this very general technique of
field-particle correlations will transform our ability to maximize the
scientific return from current, upcoming, and proposed spacecraft
missions, including the \emph{Magnetospheric Multiscale}
(\emph{MMS})\citep{Burch:2016}, \emph{Solar Probe Plus}
\citep{Fox:2015}, \emph{Turbulent Heating ObserveR} (\emph{THOR}), and
\emph{ElectroDynamics and Dissipation Interplanetary Explorer}
(\emph{EDDIE}) missions.  Further testing and refinement of this
technique will characterize its sensitivity to the noise, limited
velocity-space resolution, and limited cadence of spacecraft
measurements, as well as its ability to extract a meaningful
velocity-space signature of the collisionless damping mechanism in the
presence of the broadband spectrum of fluctuations that is
characteristic of a turbulent plasma.

This work was supported by NSF AGS-1331355, NSF CAREER Award
AGS-1054061, and DOE DE-SC0014599.

\bibliographystyle{apj}


\end{document}